\documentclass[reprint,prd,nofootinbib,superscriptaddress]{revtex4}
\usepackage{CJK}
\usepackage{amsfonts}
\usepackage{amsmath,slashed}
\usepackage{amssymb}
\usepackage[english]{babel}
\usepackage{graphicx}
\usepackage{epsfig}
\usepackage{bm}
\usepackage{verbatim}
\usepackage[utf8]{inputenc}
\usepackage{booktabs}
\usepackage{multirow}
\usepackage{subfig}
\usepackage{slashed}
\usepackage{xcolor}
\usepackage[colorlinks=true,urlcolor=red,citecolor=red]{hyperref}
\usepackage[font=small]{caption}
\usepackage{float}
\usepackage{blindtext}
\usepackage{placeins}

\newcommand{\hs}{\hspace*{0.3cm}}

\newcommand{\vs}{\vspace*{0.5cm}}
\newcommand{\be}{\begin{equation}}
\newcommand{\ee}{\end{equation}}
\newcommand{\bea}{\begin{eqnarray}}
\newcommand{\eea}{\end{eqnarray}}
\newcommand{\ben}{\begin{enumerate}}
\newcommand{\een}{\end{enumerate}}
\newcommand{\bit}{\begin{itemize}}
\newcommand{\eit}{\end{itemize}}
\newcommand{\bde}{\begin{widetext}}
\newcommand{\ede}{\end{widetext}}

\newcommand{\crn}{\nonumber \\}

\newcommand{\al}{\alpha}
\newcommand{\la}{\lambda}
\newcommand{\bet}{\beta}
\newcommand{\ga}{\gamma}
\newcommand{\va}{\varphi}
\newcommand{\om}{\omega}
\newcommand{\pa}{\partial}

\newcommand{\fr}{\frac}

\newcommand{\bc}{\begin{center}}
\newcommand{\ec}{\end{center}}

\newcommand{\de}{\delta}
\newcommand{\De}{\Delta}
\newcommand{\ep}{\epsilon}

\newcommand{\si}{\sigma}

\newcommand{\eq}{\eqref}

\newcommand{\mathsym}[1]{{}}
\newcommand{\gev}{~\mathrm{GeV}}

\definecolor{bostonuniversityred}{rgb}{0.8, 0.0, 0.0}

\newcommand{\stai}{ Subatomic Physics Research Group,
		Science and Technology Advanced Institute,\\
		Van Lang University, Ho Chi Minh City 700000, Vietnam}
\newcommand{\steh}{  Faculty of Applied Technology, School of  Technology, \\
 Van Lang University, Ho Chi Minh City 700000, Vietnam}

\begin{document}
\title{Peccei-Quinn  mechanism  and axion interactions     in the 3-3-1 model \\ 
with Cosmological Inflation}
\author{H. N. Long$^{a,b}$  }
\email{hoangngoclong@vlu.edu.vn}
\author{L. T. Hue$^{a,b}$}
\email{lethohue@vlu.edu.vn}
\affiliation{
	$^a$ \stai 
}
\affiliation{
	$^b$ \steh 
}

\date{\today }

\begin{abstract}
Based on the Peccei-Quinn (PQ) assignment, the PQ charge operator in the  3-3-1 model with Cosmological Inflation 
is constructed in  terms of diagonal generators $T_3$ and $T_8$ of the $SU(3)_L$ subgroup. The formula shows 
that difference of PQ charges  of up and down quarks is 2, i.e., $\De Q_A =2$, while for electric  charge,
are assumed to be equal, while the ($Q_A$) are opposite. PQ charge of neutral scalars equal $\pm 2$, 
while for charged scalars, it vanishes.  The couplings of axion with fermions  are presented.   
To have correct kinetic term for the axion, the PQ scale $f_{a}$ is to be VEV of the singlet 
scalar boson, namely, $f_{a}= v_\phi$. The  photon  couples  to   charged particles only, 
while in this model, the axion  does not couple to  charged  scalar/gauge bosons.  
The point is worth emphasizing that the axion has doubly derivative coupling with scalar playing the role 
of inflaton.   The new effects mainly  happen in the energy  region from $10^7 \, \gev$ to $10^{11} \, \gev$. 
The chiral effective   Lagrangian as usually provides axion mass consistent with model-independent prediction.
 
\end{abstract}


\maketitle

\section{\label{intro}Introduction}
At present, axion is one of the hottest issues in particle physics from both theoretical and experimental aspects \cite{jekim, luzio,snow,ADMX}. As shown in Refs. \cite{thoof1,thoof2}, the $\tilde{\theta}$ term: $\mathcal{L}_\theta \sim \tilde{\theta} \tilde{G}. G$   
where $\tilde{G}$ being  the dual gluon strength tensor
 is responsible for $CP$ violation.
It is well-known that the absence of $CP$ violation in strong sector of the Standard Model (SM) 
leads to upper bound of the value 
\be \tilde{\theta} \leq 10^{-10}\,.
\label{sep21}
\ee
 Question is why this value so small called
strong $CP$ problem. 

To solve the puzzle, Peccei and Quinn \cite{pq1,pq2} have represented a singlet field $\Phi (x)$ in which an axion is complex phase in polar
coordinate i.e.,  $\Phi (x) = R(x) e^{i \theta(x)}$. Here $\theta$ is responsible for neutron electric diplole moment (EDM). 
Next point in PQ recipe is to make  $\theta =0$  energetically favourable. This means that if one writes potential $V = V(\Phi)$
then it will have a minimum at desired value, for example $R = v \gg 0$. To solve the strong $CP$ problem, the potential
also favour  $\theta =0$. Setting neutron  EDM to zero, the simplest way is harmonic form of
 potential $V(\theta) \sim \theta^2 $.
 
 Combining the above discussions, one comes to redefinition of the singlet field as follows \cite{giogi}
 \be \Phi = R(x)\,  e^{i  \fr{a(x)}{f_a} x_\phi }  \,,
 \label{sep22}
\ee
where $x_\phi$ and $f_a$ are PQ charge of $\phi$ and axion decay constant, respectively. 
It is worth noting  that to satisfy the above upper, the axion decay constant
 $f_a$   must be very large:  $ f_a \geq  10^{10} \, \gev$. The factor $a/f_a $ ensures 
a solution to the strong $CP$ puzzle.

 The  PQ formalism is considered in many versions of the beyond Standard Models (BSMs).
 Among BSMs, the models based on the $\mbox{SU(3)}_C\times \mbox{SU(3)}_L\times \mbox{U(1)}_N$ (3-3-1) gauge
 group \cite{ppf1,ppf2,ppf3,ppf4,flt1,flt2,flt3,flt4,flt5,flt6} have some intriguing properties such as: i) 
Due to fact that one generation of quarks transforms differently from two other ones, the anomaly free condition 
leads to number of generation is multiple of color number. Combining with QCD asymptotic free requiring number of quarks is not  bigger than five, one gets number of generation is equal to three. By the same reason, it explains 
why top quark is so heavy (175 GeV). ii) The PQ symmetry is automatically fulfilled  in the 3-3-1 models \cite{pal}.
iii) The models provide electric charge quantization \cite{chargeq1,chargeq2}.

In the frameworks of the 3-3-1 models, the PQ formalism has been considered for almost two decades ago
 \cite{a1,a2,a3}.
However, in the above  mentioned works, the main ingredient, namely the singlet $\phi$ is expanded 
as ordinary complex scalar field, i.e., in the sum of $CP$ even and $CP$ odd components. In that cases, the axion
mixs with other $CP$ odd scalars. The axion is the pure imaginary part of $\phi$ only in the limit $v_\phi \gg v_\chi$, which is  the vacuum expectation value (VEV) responsible for breaking from $SU(3)_L$  to SM subgroup. 
In Ref. \cite{a3}, the PQ symmetry was considered for two main versions: the minimal 3-3-1 model \cite{ppf1,ppf2,ppf3,ppf4} and the version with right-handed neutrinos \cite{flt1,flt2,flt3,flt4,flt5,flt6}.

New development in this direction was done five years ago in Ref. \cite{jpf}, where the discrete symmetry $Z_{11}\times Z_2$ is imposed, and Majorana right-handed neutrinos are introduced. To provide their masses, a complex scalar transforming as a $ \mbox{SU(3)}_L$  singlet is added. As a consequence, the model also contains a heavy $CP$ even scalar with mass
in the range of $10^{10}\, \gev$ called by inflaton. However,  Ref. \cite{jpf} still contains some flawed points in the assignment of the $Z_2$ symmetry, incorrect mixing matrix of the $CP$ odd scalars, and still an absence of identification of 
the Standard Model-like Higgs boson.  The  mentioned problems
have been solved in Ref. \cite{alp331}. In the above work, using Euler rotation method,  the correct mixing matrix of $CP$ odd sector has been performed. For the $CP$ even sector, applying the Hatree-Fock approximation, the $4 \times 4$ matrix has been diagonalized. As a result, the model contains the expected inflaton with mass around $10^{11} \, \gev$, one heavy scalar with  mass at TeV scale labeled by $H_\chi$, one scalar with mass at the EW scale ($h_5$), and of course the SM-like Higgs boson ($h$).

In summary, the 3-3-1 model under consideration has some interesting properties such as: i) The existence of Majorana right-handed neutrino with  masses produced at the very high PQ scale around $10^{11}\, \gev$,
ii) The CP-even component of the singlet plays a role of the inflaton with very large mass too.
ii) Lastly, the scalar sector is quite simple and very interesting 
beside the scalar with mass 
125 GeV, there is also a partner with  mass   at the EW scale 150/96 GeV.

Note that in Refs. \cite{jpf,alp331}, the axion becomes the pure imaginary part of the singlet $\phi$, only if
$\tan \theta_\phi \approx v_\chi/v_\phi \rightarrow 0$. In addition, the axion does  not explicitly
present in the law of the PQ transformation. For example Ref. \cite{jpf}, PQ transformation of the single is $\phi
\rightarrow e^{-i2 \al X_d} \phi$, which  is quite different from that given in  Eq. \eqref{sep22}.

In this work, the singlet field $\phi$ takes complex
value everywhere, and axion is complex phase in polar coordinates. Then the axion
also appears as a phase of PQ transformations. To solve the strong $CP$ puzzles in QCD, the axion will
associate factor  $10^{-10}$ and in this case $1/f_a$ where $f_a \sim 10^{10}$ GeV.

The layout of the reminder of this work  is as follows. In section \ref{model} we briefly  present
fermion and scalar with their quantum numbers of the model.  Section \ref{pqm} is devoted to the PQ symmetry, where the PQ charge operator is constructed in terms of diagonal generators. The couplings of axion are presented in section \ref{coupling}. The anomalous and derivative axion's couplings are discussed. The axion couplings to scalar are discussed in detail in section \ref{total}. We show that to have the correct axion's kinetic form: $f_a = v_\phi$. We make outlines and conclusions in the final section - section \ref{conc}.
 The PQ transformation and its charge operator are presented in sections \ref{pqo} and \ref{formula}, respectively.

\section{Brief review of the model}
	\label{model}

 To have Dirac and Majorana mass terms for $\nu_L$ and $N_R$,  
the particles and their transforms \cite{jpf,alp331} under  $\mbox{SU(3)}_C\times \mbox{SU(3)}_L\times \mbox{U(1)}_N \times Z_2 \times Z_{11}$ group
are presented  in  Table \ref{tab1} (Reason why we have changed the label for $\mbox{U(1)}$  subgroup will be cleared
in subsection \ref{pqc}). 
	\small{
		\begin{table}[th]
			\resizebox{8cm}{!}{
				\begin{tabular}{|c|c|c|c|c|c|c|c|c|c|c|c|c|c|c|}
					\hline
					& $Q_{n L}$ & $Q_{3L}$ & $u_{a R}$ & $d_{a  R}$ & $T_{3R}$ &$ D_{n R} $ & $\psi_{aL}$ & $
					l_{aR}$ & $N_{aR} $ & $\eta$ & $\chi $ & $\rho$ & $
					\phi$  \\ \hline
					$SU(3)_C$ & $\mathbf{3}$ & $\mathbf{3}$ & $\mathbf{3}$ & $\mathbf{3}$ & $%
					\mathbf{3}$ & $\mathbf{3}$ & $\mathbf{1}$ & $\mathbf{1}$ & $\mathbf{1}$ & $%
					\mathbf{1}$ & $\mathbf{1}$ & $\mathbf{1}$ & $\mathbf{1}$   \\ \hline
					$SU(3)_L$ & $\overline{\mathbf{3}}$ & $\mathbf{3}$ & $\mathbf{1}$ & $%
					\mathbf{1}$ & $\mathbf{1}$ & $\mathbf{1}$ & $\mathbf{3}$ & $\mathbf{1}$ & $%
					\mathbf{1}$ & $\mathbf{3}$ & $\mathbf{3}$ & $\mathbf{3}$ & $\mathbf{1}$  \\ \hline
					$U(1)_N$ & $0$ & $\fr 1 3$ & $\fr 2 3$ & $-\fr 1 3$ & $\fr{2
					}{3}$ & $-\fr 1 3$ & $-\fr 1 3$ & $-1$ & $0$ & $-
					\fr 1 3$ & $-\fr 1 3$ & $\fr 2 3$ & $0$
					\\ \hline
					$Z_2$ & $1$ & $1$ & $-1$ & $-1$ & $1$ & $1$ & $1$ & $-1$ &
					$-1 $ & $-1$ & $1$
					& $-1$ & $1$ \\ \hline
$Z_{11}$ & $\om^{-1}_4$ & $\om_0$ & $\om_5$ & $\om_2$
					& $\om_3$ & $\om_4$ & $\om_1$ & $\om_3$ & $%
					\om^{-1}_5$ & $\om^{-1}_5$ & $\om^{-1}_3$ & $\om^{-1}_2$ & $\om^{-1}_1$  \\
					\hline
			\end{tabular}}
			\caption{$SU(3)_C\times SU(3)_L\times U(1)_N\times Z_{11}\times Z_2$ charge assignments of the particle content of the model.
				Here $w_k=e^{ik2\pi/11}$, $a=1,2,3$ and $\al =1,2$.}
			\label{tab1}
	\end{table}}
	
Masses of  fermions and gauge bosons request  VEVs of three triplets and one singlet
	\bea \langle \eta \rangle  & = & \fr 1{\sqrt{2}}\, \left( v_\eta , 0, 0\right)^T\,, \, \langle \chi \rangle   = \fr 1{\sqrt{2}}\left( 0 , 0, v_{\chi}\right)^T \,,
	\crn
	\langle \rho \rangle & = & \fr 1{\sqrt{2}}\, \left( 0, v_\rho, 0
	\right)^T , \,\hs
	\langle \phi \rangle  = \fr 1{\sqrt{2}}\, v_\phi\,.
	\label{eq2t}\eea

The VEV  $v_{\chi}$ triggers the spontaneous breaking of the $SU(3)_L\times U(1)_N$ gauge symmetry down to the SM electroweak gauge group, while VEVs of  $\eta$ and $\rho$ break the SM electroweak gauge group. The VEV of the singlet $v_\phi$ breaks the PQ symmetry and provides Majoran mass for $N_R$.

\subsection{Neutrino masses}
	\label{yukaw}

With the above specified particle content, the following Yukawa interactions invariant under the $\mbox{SU(3)}_C\times \mbox{SU(3)}_L\times \mbox{U(1)}_N \times Z_2 \times Z_{11}$  symmetry, arise \cite{jpf}:
With the above specified particle content, the following Yukawa couplings for quarks invariant under the  above mentioned  group,  arise \cite{jpf,alp331}:
	\bea
	-\mathcal{L}^Y_q &=&y_{1}\bar{Q}_{3L}T_{R}\chi +\sum\limits_{n,m =1}^{2}\left(	y_{2}\right) _{n,m }\bar{Q}_{n L}D_{m R}\chi ^{\ast	}\crn
	&&+ \sum\limits_{a=1}^{3}\left( y_{3}\right) _{3a}\bar{Q}_{3L}u_{aR}\eta
	+\sum\limits_{n =1}^{2}\sum\limits_{a=1}^{3}\left( y_{4}\right) _{n a}\bar{Q}_{n L}d_{aR}\eta ^{\ast }\crn
	&&+\sum\limits_{a=1}^{3}\left( y_{5}\right) _{3a}\bar{Q}_{3L}d_{aR}\rho
	+\sum\limits_{n =1}^{2}\sum\limits_{a=1}^{3}\left( y_{6}\right) _{na}\bar{Q}_{n L}u_{aR}\rho ^{\ast }
	 +\mbox{H.c.}.
	\label{s41}
	\eea
and for the lepton \cite{alp331}
\bea -\mathcal{L}^Y_l &=&	
\sum\limits_{a=1}^{3}\sum\limits_{b=1}^{3}g_{ab}%
	\bar{\psi}_{aL}l_{bR}\rho
	+\sum\limits_{a=1}^{3}\sum\limits_{b=1}^{3}\left( y_{\nu }^{D}\right) _{ab}%
	\bar{\psi}_{aL}\eta N_{bR}\crn
	&&+\sum\limits_{a=1}^{3}\sum\limits_{b=1}^{3}\left(
	y_{N}\right) _{ab}\phi \bar{N}_{aR}^{C}N_{bR}+\mbox{H.c.}.
	\label{yuka}
	\eea
The last two terms  in \eqref{yuka} contains related to neutrino mass
	\be 
	-\mathcal{L}^{Y}_{l} \supset \left( y_{\nu }^{D}\right) _{ab}%
	\bar{\psi}_{aL}\eta N_{bR}+\left(
	y_{N}\right) _{ab} \bar{N}_{aR}^{C}N_{bR}\,  \phi +\mbox{H.c.}.
	\label{type1}
	\ee 
 
	 The Dirac neutrino mass term arises from $v_\eta $, while the Majorana mass term arises from $v_\phi$. 
	 
From \eqref{type1}, it follows that 
 the light active neutrinos are generated from 	a type I seesaw mechanism mediated by right handed Majorana neutrinos. 
  Thus implying that the resulting light active neutrino mass matrix has the form \cite{alp331}
	\be
	M_{\nu}=M^D_{\nu}M^{-1}_N\left(M^D_{\nu}\right)^T,\hs  M^D_{\nu}=  \fr{v_{\eta}}{\sqrt{2}} \, y^{D}_{\nu},
	\hs  M_N=\sqrt{2}\,v_{\phi} \, y_{N}.
	\ee
The standard neutrinos get mass at eV scale, while 
$M_R \sim 10^7$ GeV \cite{jpf}.
Note that the similar script containing  axion, inflaton and seesaw exists in Ref. \cite{smash}.

\subsection{The Higgs  sector}

The full potential of the model under consideration  has the form \cite{jpf}
\bea
V_{tot} &=&   \mu_\chi^2  \chi^\dag \chi + \mu_\rho^2
\rho^\dag \rho +  \mu_\eta^2 \eta^\dag \eta + \mu^2_\phi  \phi^* \phi + \la_1 ( \chi^\dag \chi)^2 + \la_2 ( \eta^\dag
\eta)^2\crn &  & + \la_3 ( \rho^\dag \rho)^2 +
\la_4 ( \chi^\dag \chi)( \eta^\dag \eta) + \la_5 ( \chi^\dag \chi)( \rho^\dag \rho) + \la_6 ( \eta^\dag \eta)( \rho^\dag \rho)\crn
&& + \la_7 ( \chi^\dag \eta)( \eta^\dag \chi) + \la_8 ( \chi^\dag \rho)( \rho^\dag \chi) + \la_9 ( \eta^\dag \rho)( \rho^\dag \eta)\crn
&& + \la_{10} ( \phi^* \phi)^2 + \la_{11} ( \phi^* \phi)( \chi^\dag \chi) + \la_{12} ( \phi^* \phi)( \rho^\dag \rho)
\, + \la_{13} ( \phi^* \phi)( \eta^\dag \eta)\crn
&& + \left( \la_\phi\ep^{ijk} \eta_i \rho_j \chi_k \phi^*
+ \mbox{H.c.}\right) \label{poten3} \eea

For future presentation, here  we  write explicitly  the scalar fields   
	\bea \chi^T  & = &  \left( \chi^{0}_1,\chi ^{-}_2,\chi
	_{3}^{0}\right) \sim \left(1,3,-\fr 1 3\right)\, , \, \eta^T =  \left( \eta^{0}_1,\eta^{-}_2,\eta
	_{3}^{0}\right)  \sim\left(1,3,-\fr 1 3\right)\,  , 
	\crn
	\rho^T & = & \left(
	\rho^{+}_1,\rho_2 ^{0},\rho_{3}^{+}\right) \sim \left(1,3,
	\fr 2 3\right), \,\hs
	\phi =  \fr 1 2 (v_\phi + R_\phi) e^{i \fr{a}{2 f_a}x_\phi} | \sim  (1,1,0)\,,
	\label{eq2}\eea
where $ \tan \theta_{PQ} = \fr{I_\phi}{v_\phi + R_\phi} $

In this case, the axion is decoupled from the scalar sector, and this case corresponds to the limit $v_\phi \gg v_\chi $ in 
Ref. \cite{alp331}. However, it does not touch to other sectors such as $CP$ even scalars.

The VEV $v_\phi$ is responsible for the PQ symmetry breaking resulting  (see below).  
Then VEV $v_\chi$ breaks  $\mbox{SU(3)}_L\times \mbox{U(1)}_N$  to the SM group.
Two others $v_\rho, v_\eta$ are needed for the usual $ \mbox{U(1)}_Q$	symmetry. Hence, it follows  $ v_\phi \gg v_\chi \gg v_\rho, v_\eta$.	The constraint conditions at tree level 
were analyzed in Ref. \cite{a2}. From \eq{poten3}, it is quite reasonable to accept : $\la_2 \approx \la_3, \, \la_4 \approx \la_5, \, \la_7 
\approx \la_8, \,\la_{12} \approx \la_{13}$. 
From result in   Ref. \cite{padax}, 
it follows   $v_\chi \geq 10357$ GeV for $M_{Z'} \geq 4.1$ TeV.

The result of  Ref. \cite{alp331} showed that in the CP-even scalar sector, there are six fields. One massless field is part of $G_{X^0}$, another massive in TeV scale is associated to $\chi_1^0$. One heavy field with mass in the range of $10^{11}$ GeV and associated with singlet $\phi$ is identified  to inflaton $\Phi$. One SM-like Higgs boson $h$ with mass $\sim$ 125 GeV. Two remain fields include one heavy with mass at TeV scale ($H_\chi$) and another with mass at EW scale ($h_5$).

In the limit $v_\phi \gg v_\chi \gg v_\rho \gg v_\eta $, one has \cite{alp331}
		\bea
		\eta & \simeq & \left(
		\begin{array}{c}
			\fr{1}{\sqrt{2}}\left(  u +h_5 + i A_5\right) \\
			H_1^- \\
			G_{X^0}   \\
		\end{array}
		\right),\, \, \chi  \simeq 
		\left(
		\begin{array}{c}
			\chi_1^0 \\
			G_{Y^-} \\
			\fr 1{\sqrt{2}}\left( v_\chi  + H_\chi + i G_{Z'}\right) \\
		\end{array}
		\right)\, ,\,\,
		\rho \simeq 
		\left(
		\begin{array}{c}
			G_{W^+} \\
			\fr{1}{\sqrt{2}}\left(  v + h+ i G_Z\right)\\
			H_2^+ \\
		\end{array}
		\right)\,,
		\crn
		\phi & = & \fr{1}{\sqrt{2}}\left( v_\phi + \Phi\right) e^{- i \fr{ a}{\, f_a}}\, .
		\label{d25}\eea

Combination of table \ref{tab1} and   Eqs. \eq{s41}, \eq{yuka}, \eq{d25} leads to some interesting consequences: 
 Firstly, the SM-like Higgs boson $h$  has Yukawa couplings with only  SM fermions. Secondly, the pseudoscalar  $A_5$ and $H_\chi$ can have Yukawa couplings with not only exotic quarks but also SM quarks and leptons.

\section{\label{pqm} Peccei - Quinn mechanism}

\subsection{\label{pqc} PQ charge in the 3-3-1 model with Inflation}

The new format of writing PQ transforms is given in Refs. \cite{giogi,jekim,luzio} and in Refs.  \cite{gu,choi}. 
According to  Ref. \cite{a2}, and using notations in Refs. \cite{gu,choi} for an arbitrary fermion and scalar boson,  the PQ  transformations are as  follows
\bea
f & \rightarrow & f^\prime =  e^{ i  \left(\fr{x_f}{2 f_a}\right)\ga_5 a} f\,, \hs {\bar f} \rightarrow {\bar f}^\prime =  {\bar f} e^{ i  \left(\fr{x_f}{2 f_a}\right)\ga_5 a} \,, \hs
\va \rightarrow \va^\prime = e^{ i \left(\fr{x_\va}{2 f_a}\right) a} \va \,,
\label{pqr}\\
 f_L &\rightarrow& f^\prime_L  =  e^{ - i  \left(\fr{x_f}{2 f_a}\right) a} f_L\,,\hs 
{\bar f}_L \rightarrow {\bar f}^\prime_L  = {\bar f}_L  e^{  i  \left(\fr{x_f}{2 f_a}\right) a}\,,\crn
 f_R &\rightarrow & f^\prime_R  =  e^{  i  \left(\fr{x_f}{2 f_a}\right) a} f_R\,,\hs 
{\bar f}_R \rightarrow {\bar f}^\prime_R  = {\bar f}_R  e^{ - i  \left(\fr{x_f}{2 f_a}\right) a}\,,  \label{hay1}
\eea
where $x_f$ is  PQ charge of fermion and $f_a \sim 10^{11} \gev$  is  axion decay constant relating to
the scale of symmetry breaking of $U(1)_{PQ}$ global group. 
The PQ charges of fermions are given as \cite{jpf}
\be  
  x_u = x_T =  - x_d = - x_D =   x_l = - x_{l R}  = - x_\nu =  x_{\nu_R} = - x_N  \equiv R \, .  
\label{s1}
\ee
Note that the parameter $R$ in \eqref{s1} in principle,  is any non-zero integer. However, 
as seen later, within notations in \eqref{pqr},  the absolute value of $R$ should be equal to the unit, i.e.,
 $\vert R \vert = 1$. Here, we choose the sign is plus, i.e., $R = 1$.
 
 For scalars, from Yukawa couplings, it follows that  charged scalars have vanishing   PQ charge  
 since they connect {\it up and down}  particles with 
opposite values,    while electrically  neutral scalar has  PQ charge duplicate charge  (with opposite sign)  of fermion to which it provides mass,  because  it  connects to {\it both up or down} particles: 
\bea  
\eta_1^0 & \rightarrow & e^{ i \fr{a}{ f_a}} \eta_1^0\,, \hs \chi_1^0
\rightarrow e^{ i \fr{a}{ f_a}} \chi_1^0 \, ,\hs \rho_1^+
\rightarrow \rho_1^+\,,\crn 
\phi  & \rightarrow &   e^{i \fr{a}{ f_a}} \phi \,, \hs    \rho_2^0
\rightarrow e^{- i \fr{a}{ f_a}} \rho_2^0 \, ,\hs  \chi_2^-
\rightarrow \chi_2^-\,.
\label{s3}
 \eea  

The last term in \eq{poten3} shows that $\rho$  has  the same value and opposite to that of $\eta$,  $\chi$ and $\phi$. This
is  explicitly  clarified in Table \ref{tab2}.

\begin{table}[th]
\resizebox{8cm}{!}{
				\begin{tabular}{|c|c|c|c|c|c|c|c|c|c|c|c|c|c|c|c|c|c|c|}
					\hline
					&  $\, u \, $ & $\, d\, $ & $T$ & $D_\al$ & $\, l\, $ &$ \nu$ & $\nu_R$ & $
					N_R$ & $\eta_1^0 $ & $\eta_3^0$ & $\chi_1^0 $ & $\chi_3^0$ & $
					\rho^0$ &$\phi$&$\eta_2^-$&$\chi_2^-$&$\rho_1^+$&$\rho_3^+$ \\ \hline
					$U(1)_{PQ}$ & $1$ & $-1$ & $1$ & $-1$ & $
					-1$ & $+1$ & $1$ & $-1$ & $2$ & $%
					2$ & $2$ & $2$ & $-2$  & $2$  & $0$  & $0$  & $0$  & $0$ 
					\\ \hline
			\end{tabular}}
			\caption{$U(1)_{PQ}$ charge assignments of the particle content of the model.
				Here $x_F = x_{F_R} = - x_{F_L}$
				}
			\label{tab2}
	\end{table}

For singlets of right-handed fermions and scalar $\phi$, the generators $T_3$ and $T_8$ produce  the zero, but  the
general PQ charge $\mathcal{X}_{pq}$ for right handed fermions  takes opposite value of left handed partner, while
for $\phi$, as usually,  it is followed   from Yukawa couplings, i.e., $\mathcal{X}_{pq}(f_R) = - Q_A(f_L)$ and  
 $\mathcal{X}_{pq}(\phi) =  2\,, \Rightarrow Q_A(\phi) = 2$. To be noted that there exists the same situation when one deals with electric charges  of right-handed fermions.

The PQ charges of the model  particle content can be clearly drawn in Table \ref{long1}.
 
\small{	
		\begin{table}[th]
			\resizebox{14cm}{!}{
\begin{tabular}{ c | c  | c}
					\hline
			$\textrm{Particle} \vert\,  Q_A$		&  $\textrm{Particle} \vert \,  Q_A$ & $ \textrm{Particle} \vert\,  Q_A $  \\ \hline

				$  \left(
		\begin{array}{c}
			t \\
			b \\
			T \\
		\end{array}
		\right)_L$ $\left(
		\begin{array}{c}
			  1\\
			- 1\\
			 1\\
		\end{array}
		\right)\sim 3 $
		& $\left(
		\begin{array}{c}
			d_\al \\
			- u_\al  \\
			D_\al  \\
		\end{array}
		\right)_L$
$\left(
		\begin{array}{c}
			  -1\\
			 1\\
			-1\\
		\end{array}
		\right)  \sim \tilde{3}$		
		
		 & $\left(
		\begin{array}{c}
			\nu_{a} \\  
			l_{a}  \\
			(\nu^{c}_R)^a  \\
		\end{array}
		\right)_L$
		
		$\left(
		\begin{array}{c}
			  1\\
			- 1\\
			 1\\
		\end{array}
		\right)$ 
					
					\\ \hline	
						
$\left(
		\begin{array}{c}
			\chi_1^0 \\  
			\chi_2^-  \\
			\chi_3^0 \\
		\end{array}
		\right)$
		$\left(
		\begin{array}{c}
			  2\\
			0\\
			 2\\
		\end{array}
		\right)\sim 3 $
		 &$\left(
		\begin{array}{c}
			\eta_1^0 \\  
			\eta_2^-  \\
			\eta_3^0 \\
		\end{array}
		\right)$
$\left(
		\begin{array}{c}
			  2\\
			0\\
			2\\
		\end{array}
		\right)$		
		
		 &$\left(
		\begin{array}{c}
			\rho_1^+ \\  
			\rho_2^0  \\
			\rho_3^+ \\
		\end{array}
		\right)$
$\left(
		\begin{array}{c}
			  0\\
			-2\\
			0\\
		\end{array}
		\right)\sim 3$		
		
					\\ \hline
			 $ N_{aR} \sim (1,1,0, - 1)   $ & $d_{aR}, D_{\al  R}
	\sim (-1/3, 1) $ &  $t_R, T_ R
	\sim ( 2/3, -1) $ \\ \hline

 & $\phi
	\sim (1,1,0, 2) $ 	
	 \\ \hline
	\end{tabular}}
			\caption{Multiplets and $Q_A$ charges in the model}
			\label{long1}
	\end{table}}
\vs

From Tables \ref {tab2},  \ref{long1} 
and \eqref{s3}, we can formulate  PQ charge  assignment for triplets in the model under consideration
as follows (for details, see Appendix \ref{formula})
\be 
Q_A =  2 \,  T_3 - \fr 2{\sqrt{3}}\,  T_8  +  \mathcal{X}_{pq}\,.
\label{s161}
\ee

 Therefore, the PQ transformation in \eqref{hay1} and \eqref{s3} can be written in form of PQ charge operator labeled 
 by $Q_A$ as follows 
\bea 
U(1)_{PQ}:  \hs 
f & \rightarrow & f^\prime =  e^{ i  \left(\fr{a}{2 f_a}\right)\ga_5 Q_A} f\,, 
\hs {\bar f} \rightarrow {\bar f}^\prime =  {\bar f} e^{ i  \left(\fr{a}{2 f_a}\right)\ga_5 Q_A} \,, \label{s163t}\\
 f_L &\rightarrow& f^\prime_L  =  e^{ - i  \left(\fr{a}{2 f_a}\right) Q_A} f_L\,,\hs
{\bar f}_L  \rightarrow  {\bar f}^\prime_L  = {\bar f}_L  e^{  i  \left(\fr{a}{2 f_a}\right) Q_A}\,,\crn
 f_R &\rightarrow & f^\prime_R  =  e^{  i  \left(\fr{a}{2 f_a}\right) Q_A} f_R\,,\hs 
{\bar f}_R \rightarrow {\bar f}^\prime_R  = {\bar f}_R  e^{ - i  \left(\fr{a}{2 f_a}\right) Q_A}, \, \label{s162}\\
\va &  \rightarrow & \va^\prime  =  e^{ +i  \left(\fr{a}{2 f_a}\right) Q_A} \va \,. \label{s163}
\eea

 It is worth noting that in the realization  the Georgi-Kaplan-Randall (GKR) field basis,
  all fields  except the axion, transform by additive  constant.

 The following remarks are in order.
 The formula \eqref{s161} shows that difference of electric charges of up and down quarks/leptons  is 1, i.e., $\De Q =1$, 
 while for PQ charge $\De Q_A =2$. For right handed fermions, electric charges ($Q$) of left-handed and right-handed are assumed to be equal, while for ($Q_A$) is opposite. For scalar bosons,
  only neutral scalar have a  PQ  charge equal to $\pm 2$, where the sign plus for scalar in top of doublet and minus for bottom one. The charged scalar does not have a PQ charge. So the electric charge ($ Q $)  and ($ Q_A $) have  again opposite
  property. This is  illustrated in Table \ref{long3}.

\small{	
		\begin{table}[th]
			\resizebox{12cm}{!}{
				\begin{tabular}{|c|c|c|}
					\hline
					&  $U(1)_{Q} $ & $ U(1)_{Q_A} $  \\ \hline
					
					$\left(
		\begin{array}{c}
			f_u \\
			f_d \\
		\end{array}
		\right)$ & $\De Q = 1$ & $\De Q_A =  2$ 
					\\ \hline	
						
$\left(
		\begin{array}{c}
			\va^0 \\
			\va^- \\
		\end{array}
		\right)$ & $\left(
		\begin{array}{c}
			Q(\va^0) =  0 \\
			Q(\va^-) = -1 \\
		\end{array}
		\right)$ & $\left(
		\begin{array}{c}
			Q_A(\va^0) = 2 \\
			Q_A(\va^-) =  0 \\
		\end{array}
		\right)$ 
					\\ \hline
			$\textrm{Chiral fermion	} f $	&  $Q(f_R) = Q(f_L)  $ & $Q_A(f_R) = -  Q_A(f_L)  $  \\ \hline
			
\end{tabular}}
			\caption{Some characteristic  properties between $U(1)_{Q}$ and $U(1)_{Q_A}$.
				}
			\label{long3}
	\end{table}}			
			
To solve the strong $CP$ problem, the  $U(1)_{pq}$ symmetry is spontaneously breaking by  VEV 
of the singlet $\phi$ as follows
\be Q_A \langle \phi \rangle =  \fr 2{\sqrt{2}} v_\phi \neq 0\,.
\label{s153}
\ee

\subsection{\label{yuka1}Yukawa couplings and quark masses}

From Table \ref{tab2}, we see that PQ charge of neutral scalar equals twice of PQ charges of fermions receiving for which scalar provides masses, so all  Yukawa    couplings given in  \eqref{s41}    and \eqref{yuka} are
 invariant. 
 In addition, exotic quark carrying lepton number 2, while ordinary SM quarks 
 do not, so mass eigenstates of exotic quarks are their original states, while  ordinary quarks are with mass mixing. 
 Hence,  we just deal with ordinary quarks. Note that top ($t$) quark  gets mass from VEV of scalar $\eta$ while $c$ and $u$ quarks get mass from VEV of $\rho$. For down quarks, the situation is  converse. Therefore, for up quarks, one has  \cite{alp331}
\be 
	M_{u}= \fr{v_{\eta }}{\sqrt{2}} \left( 
	\begin{array}{ccc}
		\left( y_{6}\right) _{11}\fr{v_{\rho }}{v_{\eta }} & \left( y_{6}\right)
		_{12}\fr{v_{\rho }}{v_{\eta }} & \left( y_{6}\right) _{13}\fr{v_{\rho }}{%
			v_{\eta }} \\ 
		\left( y_{6}\right) _{21}\fr{v_{\rho }}{v_{\eta }} & \left( y_{6}\right)
		_{22}\fr{v_{\rho }}{v_{\eta }} & \left( y_{6}\right) _{23}\fr{v_{\rho }}{%
			v_{\eta }} \\ 
		\left( y_{3}\right) _{31} & \left( y_{3}\right) _{32} & \left( y_{3}\right)
		_{33}%
	\end{array}%
	\right) =V_{uL}\widetilde{M}_{u}V_{uR}^{\dagger }\,, \label{tass}
\ee 
with 
\be
	\widetilde{M}_{u}= \textrm{diag} \left( m_{u},m_{c},m_{t}\right)\,.
\ee
In a similar way, one gets   for down quarks
\be
	M_{d}= \fr{v_{\rho }}{\sqrt{2}} \left( 
	\begin{array}{ccc}
		\left( y_{4}\right) _{11}\fr{v_{\eta }}{v_{\rho }} & \left( y_{4}\right)
		_{12}\fr{v_{\eta }}{v_{\rho }} & \left( y_{4}\right) _{13}\fr{v_{\eta }}{%
			v_{\rho }} \\ 
		\left( y_{4}\right) _{21}\fr{v_{\eta }}{v_{\rho }} & \left( y_{4}\right)
		_{22}\fr{v_{\eta }}{v_{\rho }} & \left( y_{4}\right) _{23}\fr{v_{\eta }}{%
			v_{\rho }} \\ 
		\left( y_{5}\right) _{31} & \left( y_{5}\right) _{32} & \left( y_{5}\right)
		_{33}%
	\end{array}%
	\right) =V_{dL}\widetilde{M}_{d}V_{dR}^{\dagger } \,, \label{dass}
\ee
with
\be
	\widetilde{M}_{d}= \textrm{diag} \left( m_{d},m_{s},m_{b}\right)  \,.
\ee

In the above  matrices, all Yukawa couplings of the form $\left(y_i\right)_{ab}$ $a,b=1,2,3$; $i=3,4,5,6$ are positive
 and real. With $\al =1, 2$ and $a=\al, 3$, these couplings can be determined  by the following equations:
\be
	\left( y_{6}\right) _{n a}=\fr{\sqrt{2}}{v_{\rho }}\left( V_{uL}\widetilde{M%
	}_{u}V_{uR}^{\dagger }\right) _{n a},\hspace{1cm}\hspace{1cm}\left(
	y_{3}\right) _{3a}=\fr{\sqrt{2}}{v_{\eta }}\left( V_{uL}\widetilde{M}%
	_{u}V_{uR}^{\dagger }\right) _{3a}  \label{yu}
\ee
\be
	\left( y_{4}\right) _{n a}=\fr{\sqrt{2}}{v_{\eta }}\left( V_{dL}\widetilde{M%
	}_{d}V_{dR}^{\dagger }\right) _{n a},\hspace{1cm}\hspace{1cm}\left(
	y_{5}\right) _{3a}=\fr{\sqrt{2}}{v_{\rho }}\left( V_{dL}\widetilde{M}%
	_{d}V_{dR}^{\dagger }\right) _{3a}\,.  \label{yd}
\ee
  From (\ref{tass}) and (\ref{dass}), the diagonalized mass matrix of ordinary quarks are determined as below:
\be
	\widetilde{M}_{u,d}=\left( V_{L}^{\left( u,d\right) }\right) ^{\dagger
	}M_{u,d}V_{R}^{\left( u,d\right) }\,.  \label{MSMquarkdiag}
\ee

In general, one gets:
\bea 
	\widetilde{M}_{f} &=&\left( M_{f}\right) _{diag}=V_{fL}^{\dagger
	}M_{f}V_{fR},\hspace{1cm}\hspace{1cm} f_{\left( R,L\right) }=V_{f\left(
		R,L\right) }\widetilde{f}_{\left( R,L\right) },  \crn
	\overline{f}_{aL}\left( M_{f}\right) _{ab}f_{bR} &=&\overline{\widetilde{f}}%
	_{kL}\left( V_{fL}^{\dagger }\right) _{ka}\left( M_{f}\right) _{ab}\left(
	V_{fR}\right) _{bl}\widetilde{f}_{lR}=\overline{\widetilde{f}}_{kL}\left(
	V_{fL}^{\dagger }M_{f}V_{fR}\right) _{kl}\widetilde{f}_{lR}=\overline{%
		\widetilde{f}}_{kL}\left( \widetilde{M}_{f}\right) _{kl}\widetilde{f}%
	_{lR}=m_{f_k}\overline{\widetilde{f}}_{kL}\widetilde{f}_{kR}, \crn
	k &=&1,2,3\,. \label{physfermions}
\eea %
Here, $\widetilde{f}_{k\left( R,L\right) }$ and $f_{k\left( R,L\right) }$ ($%
k=1,2,3$) are the SM fermionic fields in the mass and interaction bases,
respectively.

From \eqref{yuka}, it follows charged leptons acquire masses from VEV of $\rho$, while the Majorana right handed
neutrinos $N_R$ get mass from $\langle \phi \rangle$. Since we focus  on the strong $CP$ problem, we leave 
lepton  part for this moment.  

Now we turn into coupling of axion with matter fields and scalar bosons

\section{Axion interactions}
\label{coupling}

\subsection{Axion gauge boson couplings}

Note that the gauge boson self-couplings in the 3-3-1 models were calculated in \cite{self1,self2}. 
To understand these couplings, let us denote $W$ to be gauge boson of the group $SU(3)$. Then the {\it electric} and {\it magnetic} fields are:
\bea W^a_{0 \nu} & \equiv & E^a_\nu\,,\hs  \nu = 1,2,3 = i \,; \,  (\textrm{space index)}; a = 1,2, \cdots 8 \crn
 B^a_k   & \equiv & \ep_{i j k} W^a_{i j} \hs i \neq j \neq k, (j,k = 1,2,3)
\label{s51}
\eea
Then the Lagrangian of gauge bosons are follows
\bea 
- 4 \, \mathcal{L}_{gauge} & \supset & a(x)  W_{ a \mu \nu} W^{a \mu \nu} = 
2 a(x)( \, W_{a 0 \nu} W^{a 0 \nu}
+ \,  W_{a i j } W^{a i j}) = 2 a(x)\,( E_{a i} E^{a i} +  B_{a k} B^{a k}) = 
2 a(x) (\textbf{E}^2 + \textbf{B}^2)\,.
\label{s52}
\eea
In the 3-3-1 models, this part contains triple  and quartic gauge boson couplings which have been presented
in Refs. \cite{self1,self2}.

Now we consider a part with the  dual tensor
\bea 
  \, \mathcal{L}_{\theta} & \supset & \fr 1 2 \ep^{\mu \nu \al \bet}
 W_{a \mu \nu} W^{a}_{ \al \bet} = 4!\,  \ep^{0 i \al \bet} W_ {a 0 i} W^{a}_{ \al \bet} =
4! \, \ep^{0 i j k } W_{ a 0 i } W^{a}_{ j k} = 4!\,  ( E_{a i} B^{a i})  = 4! \textbf{E}. \textbf{B}\label{s53t}\\
& = & 4!\,  \ep^{0 i j k }[ \pa_0 W^a_i - \pa_i W^a_0 + g f_{abc} W^b_0 W^c_i][\pa_j W^a_k - \pa_k W^a_j + g f_{ade} W^d_j W^e_k]\,.
\label{s53}
\eea
Here
\be W^a_{\mu \nu} = \pa_\mu W^a_\nu - \pa_\nu W^a_\mu + g f_{abc} W^b_\mu W^c_\nu
\label{s54}
\ee
Note that in \eqref{s54},  {\it four space-time indexes are different.}
Then $ G \tilde{G}$ is  violated by $ x_o  \rightarrow - x_0$ (T inverse) or $ x_i  \rightarrow - x_i$ (parity).
Since $CPT$ invariance, one has $CP$ violation of $CT$ violation.

According  to Ref. \cite{pq2}, for Lagrangian below 
\bea 
\mathcal{L} = - \fr 1 4 F_{\mu \nu}^a  F^{a \mu \nu} + i  {\bar \psi} \ga_\mu D^\mu \psi + L_{Yukawa} - V(\phi)\,,
\eea
with transformation
\be \psi \rightarrow e^{i \si \ga_5} \psi\,,\hs  {\bar \psi} \rightarrow {\bar \psi}  e^{i \si \ga_5} \,
  \hs \phi \rightarrow  e^{-2 i \si } \phi\,.
\label{s111}
\ee
then one has 
\be \pa^\mu J_{\mu}^5  = \fr{g^2}{16 \pi^2}  F_{\mu \nu}^a \tilde{F}^{a \mu \nu}
\label{s112}
\ee

The the 3-3-1 model with three gauge subgroups, the anomalous couplings have the form   \cite{agauge}
\be
\mathcal{L}_{a g} = c_{G G} \fr{\al_s}{4 \pi} \fr{a}{2 f_a} G \tilde{G} + 
c_{WW}\fr{\al_2}{4 \pi} \fr{a}{2 f_a} W \tilde{W}
+ c_{BB}\fr{\al_1}{4 \pi} \fr{a}{2 f_a} B \tilde{B}\,.
\label{s46}
\ee

Using Eq. \eqref{s54} and replace $W^a$ by physical fields in given  in 
\eqref{hue1},
one will have coupling of axion ($a$) with gauge bosons
with {\it four external indexes (space-time) different.}

\subsection{Axion-fermion couplings}
	\label{Fera}
Let us start from kinematic term of fermion as follows
\bea 
\mathcal{L}^0_f & = & \fr i 2 \bar{f}  \ga^\mu \stackrel{\leftrightarrow}{\pa_\mu}  f  =  \fr i 2 (\bar{f}  \ga^\mu \pa_\mu f - \pa_\mu \bar{f} \ga^\mu f)\,.
\label{h1}
\eea
For the gauge version, where ordinary derivative replaced by covariant one,  we have
\bea 
\mathcal{L}_f & = & \fr i 2 (\bar{f}  \ga^\mu D_\mu f - D_\mu \bar{f} \ga^\mu f)\crn
& = & \fr i 2 [ \bar{f}  \ga^\mu (\pa_\mu - i P_\mu^f ) f - (\pa_\mu \bar{f} + i \bar{f} P_\mu^f )\ga^\mu 
f]\crn
& = & \fr i 2 \bar{f}  \ga^\mu \stackrel{\leftrightarrow}{\pa_\mu}  f  + \bar{f} P_\mu^f \ga^\mu 
f  \equiv  I + II
\label{h2}
\eea
Here $P_\mu = g_S G_\mu - g A_\mu - g^\prime B_\mu$ and  $G_\mu, A_\mu, B_\mu$ are matrices of gluon, $SU(3)_L$ and $U(1)_X$ bosons, respectively.

We next deal with the first term 
\bea 
I & = & \fr i 2 \bar{f}^\prime  \ga^\mu \stackrel{\leftrightarrow}{\pa_\mu}   f^\prime = \fr i 2 \bar{f} e^{  i \left(\fr{a}{2 f_a}\right)\ga_5 Q_A}  \ga^\mu \stackrel{\leftrightarrow}{\pa_\mu}   e^{ i \left(\fr{a}{2 f_a}\right) \ga_5 Q_A} f  \crn
& = & \fr i 2 \left[\bar{f}  e^{  i \left(\fr{x_f}{2 f_a}\right)\ga_5 a }  \ga^\mu  \pa_\mu ( e^{  i \left(\fr{x_f}{2 f_a}\right)\ga_5 a} f)  - \pa_\mu(\bar{f}  e^{ i \left(\fr{x_f}{2 f_a}\right) \ga_5 a} ) \ga^\mu    e^{ i \left(\fr{x_f}{2 f_a}\right) \ga_5 a } f \right]\crn
& = & \fr 1 2 \left\{ \bar{f}  e^{ i \left(\fr{x_f}{2 f_a}\right) \ga_5 a } i \ga^\mu \left[  i \left(\fr{x_f}{2 f_a}\right)  e^{ i \left(\fr{x_f}{2 f_a}\right) \ga_5 a } \ga_5 \pa_\mu a \,  f + e^{ i \left(\fr{x_f}{2 f_a}\right) \ga_5 a } \pa_\mu f \right]
\right. \crn 
& - &\left. \left[ (\pa_\mu \bar{f})  e^{  i \left(\fr{x_f}{2 f_a}\right) \ga_5 a} + i \bar{f} \left(\fr{x_f}{2 f_a}\right)   e^{  i \left(\fr{x_f}{2 f_a}\right) \ga_5 a } \ga_5 \pa_\mu a \right] i  \ga^\mu    e^{ i \left(\fr{x_f}{2 f_a}\right) \ga_5 a } f \right\}\crn
& = &  \fr i 2 \bar{f}  \ga^\mu \stackrel{\leftrightarrow}{\pa_\mu}  f  -\fr{ i^2}{2}\times   2\left(\fr{ x_f}{2 f_a}\right) (\pa_\mu a) \bar{f} \ga_5 \ga^\mu  \, f \,.
\crn
& = &  \fr i 2 \bar{f}  \ga^\mu \stackrel{\leftrightarrow}{\pa_\mu}  f  + \left(\fr{ x_f}{2 f_a}\right) (\pa_\mu a) \bar{f} \ga_5 \ga^\mu  \, f \,.
\label{h3}
\eea
The first term  in \eqref{h3} is kinematic term of fermion: $I \supset \fr 1 2 
\bar{f} i \ga^\mu \stackrel{\leftrightarrow}{\pa_\mu} f $.
The second term is th  famous derivative coupling of axion to fermion:
\bea
\mathcal{L}_{(f-a)} & = & \left(\fr{x_f}{2 f_a}\right) \pa_\mu a \bar{f} \ga_5 \ga^\mu  \, f \,.
 \label{s7}
\eea
For up quarks with $x_f=1$, one has
\be 
\mathcal{L}_{(u-a)} =   \left(\fr{1}{ 2f_a}\right) \pa_\mu \, a \, \bar{u} \ga^\mu  \ga_5 u\,.
\label{s8}
\ee

We  continue to the second term
\bea
II^\prime & = & \bar{f} e^{- i \left(\fr{x_f}{2 f_a}\right)\ga_5  a}  \ga^\mu  P_\mu^f e^{- i \left(\fr{x_f}{2 f_a}\right)\ga_5 a} f =  \bar{f} \ga^\mu  P_\mu^f f = II
\eea
This means that there are not new couplings, and this term just gives couplings between gauge boson  and two fermions.

Thus the axion - fermion  derivative  couplings  have the form
\bea
\mathcal{L}_ {(f-a)} & = &  + \left(\fr{1}{ \, f_a}\right) \pa_\mu \, a \,\left[ \bar{d}\,  {\bf c}_{d} \,\ga^\mu 
 \ga_5 d + \bar{u}\, {\bf c}_{u}\,\ga^\mu  \ga_5 u   + \bar{T}\, {\bf c}_{T}\,  \ga^\mu  
 \ga_5 T + \bar{D}_\al \, {\bf c}_{D_\al} \, \ga^\mu   \ga_5 D_\al \right. \label{s8}\\
 && + \left. \bar{l}\,  {\bf c}_{l}\ga^\mu \, \ga_5 l  + I_{\nu}\bar{\nu}_a \, 
 {\bf c}_{\nu}\, \ga^\mu  \ga_5 \nu_a + \fr 1 2  \bar{N}_{a}\, {\bf c}_{N_a} \, \ga^\mu  P_R  N_{a}  
 \right]\,.
 \label{s9}
\eea
In  \eqref{s9}, for the  coefficients  (${\bf c}_{f}, f= d,u,\cdots N_R$), one has to  count  the number of color, flavor indexes and PQ charge $Q_A(f)$.  
In the model under consideration, these coefficients have the form
 \bc
$ \textbf{c}_F$ =  $\left\{%
\begin{array}{ll} + 9 \,  \,  \hbox{for f = up quarks}
 &  3(\textrm{color number}). 3 (\textrm{flavor number}) , \\
 -  9 \,  \,  \hbox{for f = down quarks}
 &  3(\textrm{color number}). 3 (\textrm{flavor number}) , \\
  +  3 \,  \,  \hbox{for f = T exotic  quark}
 &  3(\textrm{color number}). 1 (\textrm{flavor number}) , \\
  - 6 \,  \,  \hbox{for f = $ D_\al$ exotic  quarks }
 &  3(\textrm{color number}). 2 (\textrm{flavor number}) , \\
  - 3 \,  \,  \hbox{for f = charged leptons }
 &  1(\textrm{color number}). 3 (\textrm{flavor number}) , \\
+   3 \,  \,  \hbox{for f = neutrinos }
 &  1(\textrm{color number}). 3 (\textrm{flavor number}) , \\
 -  3 \,  \,  \hbox{for f = RH Majorana neutrino }
 &  1(\textrm{color number}). 3 (\textrm{flavor number}) , \\
\end{array}%
\right.$ \ec
Our result  coincides with that in Ref. \cite{neu}. 

Note that the above  derivative couplings  are completely  different from usual Yukawa ones in Ref. \cite{alp331}.

\subsection{Axion-scalar couplings}
	\label{Scala}
	
We  continue to scalar sector.  For complex scalar $\va$, its  Lagrangian is
\bea \mathcal{L}_\va & = & (D^\mu \va)^\dag D_\mu \va  = [(\pa_\mu - i P^\va_\mu )\va]^\dag (\pa^\mu - i
 P^{\va \mu })\va\crn
& = & \pa^\mu \va^\dag \pa_\mu \va  - i \pa_\mu \va^\dag 
P^{\va \mu } \va + i \va^\dag  P^{\va \mu} \pa_\mu \va 
+ \va^\dag   P^\va_\mu   P^{\va \mu} \va \equiv  \textbf{A1} +  \textbf{A2} + \textbf{A3}  \,.
 \label{s10}
\eea
Within PQ transformation in Eq. \eqref{s3}, one has
\bea 
\textbf{A1} & = &   \pa^\mu \left(\va^\dag e^{  i \left(\fr{a}{2 f_a}\right) Q_A}\right)  \pa_\mu \left( e^{- i
 \left(\fr{a}{2 f_a}\right) Q_A}\va\right) =  \pa^\mu \left(\va^\dag e^{  i \left(\fr{x_\va}{2 f_a}\right) a}\right)  \pa_\mu \left( e^{ - i \left(\fr{x_\va}{2 f_a}\right) a}\va\right)\crn
& = &  \left[\pa_\mu \va^\dag . e^{ i \left(\fr{x_\va}{2 f_a}\right) a} +
\va^\dag e^{ i \left(\fr{x_\va}{2 f_a}\right) a} ( i)\left(\fr{x_\va}{2 f_a}\right) \pa_\mu a 
\right] \left[ i \left(\fr{x_\va}{2 f_a}\right) e^{ - i \left(\fr{x_\va}{2 f_a}\right) a} \pa^\mu a\,  \va +  e^{ - i \left(\fr{x_\va}{2 f_a}\right) a} \pa^\mu \va \right]\crn
& = & \pa_\mu \va^\dag \pa^\mu \va + \left(\fr{x_\va}{2 f_a}\right)^2  \pa_\mu a \pa^\mu a \,  \va^\dag \va - i \left(\fr{x_\va}{2 f_a}\right) \pa_\mu a (\pa^\mu \, \va^\dag \va - \va^\dag \pa^\mu \va)\crn
& = & \textrm{ kinetic term of} \hs \va + \left(\fr{x_\va}{2 f_a}\right)^2  \pa_\mu a \pa^\mu a \,  \va^\dag \va -  i \left(\fr{x_\va}{2 f_a}\right) \pa_\mu a (\pa^\mu \va^\dag \va - \va^\dag \pa^\mu \va)
\label{s11}
\eea
Applying Eq. \eqref{s11} for the model under consideration, one gets  the first term in \eqref{s11} is kinematic for scalars. There are quartic couplings of two axions with two {\it neutral} scalars and triple couplings of one axion to two neutral scalars. Namely
\bea 
\textbf{A1} & = & \textrm{ kinetic term of} \hs H + \left(\fr{1}{ f_a}\right)^2  \pa_\mu a \pa^\mu a \left(\sum_{H= \eta_1^0, \eta_3^0,
\rho_2^0}^{ \chi_1^0, \chi_3^0, \phi} H^* H\right) \crn
& - & i \left(\fr{x_\va}{2 f_a}\right) \pa^\mu a \sum_{D=\eta,\chi}^{K=\rho^0 \phi^0} \left[D^*\stackrel{\leftrightarrow}{\pa_\mu} D - K^*\stackrel{\leftrightarrow}{\pa_\mu} K\right]
\label{s12}
\eea

The explicit results are
\bea 
\textbf{A1331}& \equiv & \textbf{A1}^{331}  =   \textrm{ kinetic term of} \hs \textrm{scalars} + \left(\fr{1}{ f_a}\right)^2  \pa_\mu a \pa^\mu a 
\left( \eta_1^{0*} \eta_1^0 +  \eta_3^{0*} \eta_3^0 + \chi_1^{0*} \chi_1^0 +  \chi_3^{0*} \chi_3^0 + \rho_2^{0*} \rho_2^0 + \phi^{0*} \phi^0\right)\crn
&-& i \left(\fr{1}{ f_a}\right) \pa^\mu a 
\left[\eta_1^{0*}\stackrel{\leftrightarrow}{\pa_\mu} \eta_1^{0} + \eta_3^{0*}\stackrel{\leftrightarrow}{\pa_\mu} \eta_3^{0} + \chi_1^{0*}\stackrel{\leftrightarrow}{\pa_\mu} \chi_1^{0} + \chi_3^{0*}\stackrel{\leftrightarrow}{\pa_\mu} \chi_3^{0}
- \rho_2^{0*}\stackrel{\leftrightarrow}{\pa_\mu} \rho_2^{0}  - \phi^{0*}\stackrel{\leftrightarrow}{\pa_\mu} \phi^{0}  \right]. 
\label{s13}
\eea 

Next, for the second term in Eq. \eqref{s10}
\bea 
\textbf{A2} & = &  - i \pa_\mu \left[\va^\dag e^{ i \left(\fr{x_\va}{2 f_a}\right) a} \right]
P^{\va \mu } e^{- i \left(\fr{x_\va}{2 f_a}\right) a}  \va + i \va^\dag e^{ i \left(\fr{x_\va}{2 f_a}\right) a}  P^{\va \mu} \pa_\mu \left[ e^{- i \left(\fr{x_\va}{2 f_a}\right) a} \va\right]\crn
& = &  - i \left[(\pa_\mu \va^\dag )e^{ + i \left(\fr{x_\va}{2 f_a}\right) a} + i \va^\dag 
\left(\fr{x_\va}{2 f_a}\right) \pa_\mu a e^{  i \left(\fr{x_\va}{2 f_a}\right) a}\right]
P^{\va \mu } e^{- i \left(\fr{x_\va}{2 f_a}\right) a}  \va \crn
& + & i \va^\dag e^{ i \left(\fr{x_\va}{2 f_a}\right) a}  P^{\va \mu}\left[
- i \left(\fr{x_\va}{2 f_a}\right)  \pa_\mu  a   e^{- i \left(\fr{x_\va}{2 f_a}\right) a} \va + e^{- i \left(\fr{x_\va}{2 f_a}\right) a} \pa_\mu \va \right]\crn
& = &  - i \left[\pa^\mu \va^\dag  + i \va^\dag 
\left(\fr{x_\va}{2 f_a}\right) \pa^\mu a \right]
P_\mu^{\va }  \va \crn
& - & i \va^\dag   P_\mu^{\va }\left[
 i \left(\fr{x_\va}{2 f_a}\right)  \pa^\mu  a  \,  \va -  \pa^\mu \va \right]
\crn
& = & - i \left(\pa^\mu \va^\dag P_\mu^{\va }  \va  -  \va^\dag   P_\mu^{\va }\pa^\mu \va \right)
+  2 \left(\fr{x_\va}{2 f_a}\right) (\pa^\mu a\,)  \left[ \va^\dag P_\mu^{\va } \va\right]. 
\label{s14}
\eea
The first term in Eq. \eqref{s14} is ordinary one in the 3-3-1 model (without axion's participation). The second term is quartic couplings of axion, gauge  and two scalar bosons.

There are no new coupling in \textbf{A3}, or other word speaking, \textbf{A3} contains ordinary couplings of scalars to gauge bosons.

In summary, new couplings of axion with neutral  scalars are (including operators of up to dimension 5)

\bea
\mathcal{L}_{(a S)} & = & \left(\fr{1}{ f_a}\right)^2  \pa_\mu a \pa^\mu a \left(\sum_{H= \eta_1^0, \eta_3^0,
\rho_2^0}^{ \chi_1^0, \chi_3^0, \phi} H^* H\right)\crn
&- & i \left(\fr{x_\va}{ 2 f_a}\right) \left( \pa^\mu a\,\right)  \sum_{D=\eta,\chi}^{K=\rho^0 \phi^0} \left[D^*\stackrel{\leftrightarrow}{\pa_\mu} D - K^*\stackrel{\leftrightarrow}{\pa_\mu} K\right]
\crn
& + & 2 \left(\fr{x_\va}{2 f_a}\right) \left( \pa^\mu a\,\right)  H^\dag P_\mu^{H } H
  \crn&\equiv& \mathcal{L}(aa HH) + \mathcal{L}(a HH) + \mathcal{L}(a G HH)\,,\label{s15}
\eea
where $G$ is labeled for gauge bosons.
 For more detail of axion couplings, the reader is referred to  Refs. \cite{choi,agauge,luzio,agu}.
 
\subsection{Interactions of  axion to scalar and gauge  bosons} 
\label{sax}

We proceed now explicit form of the above terms.
Note that in Eq. \eqref{s15}, there are not only couplings of axion to  scalar bosons, but also to gauge bosons. 
Let us consider quartic  couplings
\bea 
\mathcal{L}(aaHH) & = & \left(\fr{1}{f_a}\right)^2  \pa_\mu a \pa^\mu a \left(\sum_{H= \eta_1^0, \eta_3^0,
\rho_2^0}^{ \chi_1^0, \chi_3^0, \phi} H^* H\right)\crn
& = &  \left(\fr{1}{f_a}\right)^2  \pa_\mu a \pa^\mu a 
\left( \eta_1^{0*} \eta_1^0 +  \eta_3^{0*} \eta_3^0 + \chi_1^{0*} \chi_1^0 +  \chi_3^{0*} \chi_3^0 + \rho_2^{0*} \rho_2^0 + \phi^{*} \phi\right)\crn
&=& \fr 1 2\left(\fr{1}{f_a}\right)^2  \pa_\mu a \pa^\mu a \left\{ \fr{}{}\left[(v_\eta + R^1_\eta)^2 +  (I^1_\eta)^2 + (v_\chi 
+ R^3_\chi)^2 +  (I^3_\chi)^2 + (v_\rho 
+ R_\rho)^2 +  (I_\rho)^2 
 +  (v_\phi + R_\phi)^2  \right]\right. \crn
 &  &\left.  \qquad \qquad \qquad  \qquad +2 ( \eta_3^{0*} \eta_3^0 +  \chi_1^{0*} \chi_1^0) \fr{}{}\right\}
\crn
 & = &  \fr 1 2 \left(\fr{1}{f_a}\right)^2  \pa_\mu a \pa^\mu a \left[ \fr{}{}\,  (v_\eta^2 + v_\rho^2 + v_\chi^2 +v_\phi^2) \right. \label{s19ta}
 \\
 && \qquad \qquad \qquad \qquad + 2 \left( v_\eta R^1_\eta +  v_\rho R_\rho +  v_\chi R^3_\chi + v_\phi R_\phi\right) \crn
 && \qquad \qquad \qquad \qquad + \left.  (R^1_\eta)^2 +  (I^1_\eta)^2 + (R_\rho)^2 +  (I_\rho)^2 + (R^3_\chi)^2 +  (I^3_\chi)^2 +(R_\phi)^2  +2( \eta_3^{0*} \eta_3^0 +  \chi_1^{0*} \chi_1^0)  \fr{}{}\right]. 
\eea

From Eq. \eqref{s19ta}, it follows that  to get a right form of axion  kinetic term, the following condition is  
\be f^2_{a} = v_\eta^2 + v_\rho^2 + v_\chi^2 +v_\phi^2 \,.
 \label{h5}
\ee

Thus one gets  finally
\bea 
\mathcal{L}(aaHH) & = &  \fr 1 2  \pa_\mu a \pa^\mu a
 + \left(\fr{1}{f_a}\right)^2  \pa_\mu a \pa^\mu a \, \left\{  v_\eta R^1_\eta +  v_\rho R_\rho +  v_\chi R^3_\chi + v_\phi R_\phi +     \eta_3^{0*} \eta_3^0 + \chi_1^{0*} \chi_1^0 \right. \label{s19}\\
 && + \fr 1 2 \left. \left[ (R^1_\eta)^2 +  (I^1_\eta)^2 + (R_\rho)^2 +  (I_\rho)^2 + (R^3_\chi)^2 +  (I^3_\chi)^2 +(R_\phi)^2 \right]   \right\}
 \label{s19t}
\eea
In Eq. \eqref{s19}, there exist triple couplings of two axions to one CP-even scalar. One of them with the
strength of  coupling is $\pa_\mu a \pa^\mu a \,   R_\phi = v_\phi (f_a)^{-2} \sim  (f_a)^{-1}$ - the coupling of two axions to inflaton. 
We should add this interaction  for the future study, 
 since  $\propto (f_a)^{-1}$, while terms in the second line can be neglected.
 
We  next consider triple couplings of an axion to two scalars.
\bea 
\mathcal{L}(aHH) & = & - i \left(\fr{1}{f_a}\right) \pa^\mu a 
\left(\eta_1^{0*}\stackrel{\leftrightarrow}{\pa_\mu} \eta_1^{0} + 
\eta_3^{0*}\stackrel{\leftrightarrow}{\pa_\mu} \eta_3^{0} + 
\chi_1^{0*}\stackrel{\leftrightarrow}{\pa_\mu} \chi_1^{0} + \chi_3^{0*}
\stackrel{\leftrightarrow}{\pa_\mu} \chi_3^{0}\right.\crn
&-&   \left. \rho_2^{0*}\stackrel{\leftrightarrow}{\pa_\mu} \rho_2^{0}  -
 \phi^{0*}\stackrel{\leftrightarrow}{\pa_\mu} \phi^{0}  \right)\crn
& = & - i \left(\fr{1}{f_a}\right) \pa^\mu a \left( - i v_\eta \pa_\mu I^1_\eta + i I^1_\eta \pa_\mu R^1_\eta 
-i R^1_\eta \pa_\mu I^1_\eta   - i v_\chi \pa_\mu I^3_\chi + i I^3_\chi \pa_\mu R^3_\chi
 -i R^3_\chi \pa_\mu I^3_\chi \right. \crn
& &  - i v_\rho \pa_\mu I_\rho + i I_\rho \pa_\mu R_\rho
-i R_\rho \pa_\mu I_\rho \crn
&&\left.  + \eta_3^{0*}\stackrel{\leftrightarrow}{\pa_\mu}
 \eta_3^{0} + \chi_1^{0*}\stackrel{\leftrightarrow}{\pa_\mu} \chi_1^{0}\right)\crn
& = &-    \left(\fr{1}{f_a}\right) \pa^\mu a \, \left(v_\eta \pa_\mu I^1_\eta +  v_\chi \pa_\mu I^3_\chi
+ v_\rho \pa_\mu I_\rho \right)  \crn
&  & -  i \left(\fr{1}{f_a}\right) \pa^\mu a \left(  i I^1_\eta \pa_\mu R^1_\eta  
+ i I_\rho \pa_\mu R_\rho + i I^3_\chi \pa_\mu R^3_\chi 
\right. \crn
& & -i  R^1_\eta \pa_\mu I^1_\eta     -i R^3_\chi \pa_\mu I^3_\chi  -i R_\rho \pa_\mu I_\rho  \crn
&&\left.  + \eta_3^{0*}\stackrel{\leftrightarrow}{\pa_\mu} \eta_3^{0}
+ \chi_1^{0*}\stackrel{\leftrightarrow}{\pa_\mu} \chi_1^{0}\right)\crn
& = & \mathcal{L}_{(a A)} + \mathcal{L}_{(a SS)}\,,
\label{s20}
\eea
where the mixing of axion with CP-odd scalars is determined as
\be
 \mathcal{L}_{(a A)} = - \left(\fr{1}{f_a}\right) \pa^\mu a \, \left(v_\eta \pa_\mu I^1_\eta +  v_\chi \pa_\mu I^3_\chi + v_\rho \pa_\mu I_\rho \right)\,.
\label{s21a}
\ee
The  terms in Eq. \eqref{s21a} are some kind of mixing between axion and Goldstone bosons eaten by neutral gauge bosons. 
This  unwanted mixing could be rotated away by making $U(1)$ rotation \cite{giogi}.
Hence one has coupling of axion to two scalar fields given below
\bea
 \mathcal{L}_{(a SS)} & = &- i \left(\fr{1}{f_a}\right) \pa^\mu a \left(  i I^1_\eta \pa_\mu R^1_\eta  + i I_\rho \pa_\mu R_\rho + i I^3_\chi \pa_\mu R^3_\chi 
\right. \crn
& &-i  R^1_\eta \pa_\mu I^1_\eta     -i R^3_\chi \pa_\mu I^3_\chi  -i R_\rho \pa_\mu I_\rho  \crn
&&\left.  + \eta_3^{0*}\stackrel{\leftrightarrow}{\pa_\mu} \eta_3^{0} + \chi_1^{0*}\stackrel{\leftrightarrow}{\pa_\mu} \chi_1^{0}\right)
\label{s22}
\eea
We close this part by considering 
\bea 
\mathcal{L}(aGHH) & = &  2 \left(\fr{x_\va}{2 f_a}\right) \pa^\mu a\,  \va^\dag P_\mu^{\va } \va\crn
 & = &  2 \left(\fr{x_\va}{2 f_a}\right) \pa^\mu a\, [ \eta^\dag P_\mu^{\eta } \eta +
  \chi^\dag P_\mu^{\chi } \chi +  \rho^\dag P_\mu^{\rho } \rho]\crn
 & = &  2 \left(\fr{g}{f_a}\right) \pa^\mu a \left\{ \left(W_{3 \mu}+ \fr 1{\sqrt{3}} W_{8 \mu}-\fr 1 3 t \sqrt{\fr 2 3} B_\mu \right)  v_\eta^2    - \left(W_{3 \mu} -  \fr 1{\sqrt{3}} W_{8 \mu}- \fr 2 3 t \sqrt{\fr 2 3} B_\mu \right) v_\rho^2 \right.  \crn
  && - \left.\left( \fr 2{\sqrt{3}} W_{8 \mu} + \fr 1 3 t \sqrt{\fr 2 3} B_\mu \right) v_\chi^2
 \right\}\crn
 && + 2 \left(\fr{g}{f_a}\right) \pa^\mu a\,\left\{\left(W_{3 \mu}+ \fr 1{\sqrt{3}} W_{8 \mu}-\fr 1 3 t \sqrt{\fr 2 3} B_\mu \right)[ 2 v_\eta  R^1_\eta +  (R^1_\eta )^2 +  (I^1_\eta)^2]\right. \crn
 && + \left(-W_{3 \mu}+ \fr 1{\sqrt{3}} W_{8 \mu}+\fr 2 3 t \sqrt{\fr 2 3} B_\mu \right)[ 2 v_\rho  R^2_\rho +  (R_\rho )^2 +  (I_\rho)^2] \crn
 && + \left(- \fr 2{\sqrt{3}} W_{8 \mu}- \fr 1 3 t \sqrt{\fr 2 3} B_\mu \right)[ 2 v_\chi 
  R^3_\chi +  (R^3_\chi )^2 +  (I^3_\chi)^2] \crn
  && + \left. \left(- \fr 2{\sqrt{3}} W_{8 \mu}-\fr 1 3 t \sqrt{\fr 2 3} B_\mu \right) \eta_3^{0*}\eta_3^0  
  +  \left(W_{3 \mu}+ \fr 1{\sqrt{3}} W_{8 \mu}-\fr 1 3 t \sqrt{\fr 2 3} B_\mu \right) \chi_1^{0*}\chi_1^0
   \right\}\crn
   & = & \mathcal{L}_{(a W)} + \mathcal{L}_{aGHH}
\label{s21b}
\eea
where the first term in Eq. \eqref{s21b} is term mix axion with weak gauge bosons
\bea 
 \mathcal{L}_{(a W)} & = &
 2 \left(\fr{g}{f_a}\right) \pa^\mu a \left[ \left( \fr{Z_{\mu}}{c_W} + Z'_{\mu} \fr{1 -t_W^2}{\sqrt{3- t_W^2}}\right) v_\eta^2     +\left(\fr{Z'_{2\mu }}{c_W^2 \sqrt{3-t_W^2}}-\fr{Z_{\mu }}{c_W}\right) v_\rho^2
  + \left(-\fr{2 Z'_{\mu }}{\sqrt{3-t_W^2}}\right)   v_\chi^2
 \right]
 \crn
 & = &  vev^2. \pa_\mu a\,  Z^\mu   \,,
\label{s21t}
\eea
and
\bea 
 \mathcal{L}_{aGHH} & = & 2 \left(\fr{g}{f_a}\right) \pa^\mu a\,\left[ \left( \fr{Z_{\mu}}{c_W} + Z'_{\mu} \fr{1 -t_W^2}{\sqrt{3- t_W^2}}\right)[ 2 v_\eta  R^1_\eta +  (R^1_\eta )^2 +  (I^1_\eta)^2]\right. \crn
   &&  + \left(\fr{Z'_{2\mu }}{c_W^2 \sqrt{3-t_W^2}}-\fr{Z_{\mu }}{c_W}\right)[ 2 v_\rho  R^2_\rho +  (R_\rho )^2 +  (I_\rho)^2] \crn
   && + \left(-\fr{2 Z'_{\mu }}{\sqrt{3-t_W^2}}\right) [ 2 v_\chi 
   R^3_\chi +  (R^3_\chi )^2 +  (I^3_\chi)^2] \crn
   && + \left. \left(-\fr{2 Z'_{\mu }}{\sqrt{3-t_W^2}}\right) \eta_3^{0*}\eta_3^0  
   +   \left( \fr{Z_{\mu}}{c_W} + Z'_{\mu} \fr{1 -t_W^2}{\sqrt{3- t_W^2}}\right) \chi_1^{0*}\chi_1^0
   \right]
 \crn & = &  \textrm{Couplings}(aG
   HH),
\label{s21t2}
\eea 
where  $t=\fr{3 \sqrt{2} t_W}{\sqrt{3-t_W^2}}$.  We also use  the limit $v^2_{\chi}\gg v^2$, which  result in  the following relations between physical basis $(A_{\mu},Z_{\mu}, Z'_{\mu})$ and the original gauge boson states \cite{flt4}: 
\begin{align} \label{hue1}
 W_{3\mu} =& A_{\mu } s_W+c_W Z_{\mu },
\crn W_{8\mu}=&\left( -A_{\mu } c_W + s_W Z_{\mu } \right) \fr{ t_W}{\sqrt{3}}+\fr{\sqrt{3-t_W^2} Z'_{\mu }}{\sqrt{3}},
\crn B_{\mu}=& -\left( -A_{\mu } c_W + s_W Z_{\mu } \right) \fr{\sqrt{3-t_W^2}}{\sqrt{3}} +\fr{t_W Z'_{\mu }}{\sqrt{3}},
\end{align}

In Eq. \eqref{s21t2}, there are couplings, up to dimension 5, between  axion, gauge boson with one scalar or two scalars.

To close this section, it is worth noting that unlike photon,  the axion  does not couple to charged scalar and gauge bosons.

\section{\label{total}Total axion Lagrangian}
The total part concerned to axion is given below
\bea
\mathcal{L}_a & = & \fr 1 2 \pa_\mu a \pa^\mu a - \fr 1 2 m_{a o }^2 \, a^2 
 +  c_{G G} \fr{\al_s}{4 \pi} \fr{a}{2 f_a} G \tilde{G} + 
c_{WW}\fr{\al_2}{4 \pi} \fr{a}{2 f_a} W \tilde{W}
+ c_{BB}\fr{\al_1}{4 \pi} \fr{a}{2 f_a} B \tilde{B}
\label{s181}\\
& + &  \fr{\pa^\mu a}{2 f_a} \left(\sum_{f=u,d,T,D}^{l,\nu}  \bar{\psi}_f c_f \ga_\mu \ga_5 \psi  + \fr 1 2  \bar{N}_{a}\, {\bf c}_{N_a} \, \ga^\mu  P_R  N_{a} \right) -  \left(\bar{q}_L\,  M_a\,  q_R + \mbox{H.c.}\right)\label{s182}\\
& + &  \left(\fr{1}{ f_a}\right)^2  \pa_\mu a \pa^\mu a \left(\sum_{H= \eta_1^0, \eta_3^0,
\rho_2^0}^{ \chi_1^0, \chi_3^0, \phi} H^* H\right)\hs \hs ;  \label{s183}\\
&-& i \left(\fr{x_\va}{ 2 f_a}\right) \pa^\mu a \sum_{D=\eta,\chi}^{K=\rho^0 \phi^0} \left[D^*\stackrel{\leftrightarrow}{\pa_\mu} D - K^*\stackrel{\leftrightarrow}{\pa_\mu} K\right]
\label{s184}\\
& + & 2 \left(\fr{x_\va}{2 f_a}\right)  \pa^\mu a\,\sum_{H= \eta_1^0, \eta_3^0,
\rho_2^0}^{ \chi_1^0, \chi_3^0, \phi}  H^\dag P_\mu^{H } H
\,,\label{s185}
\eea   
where $H^* H= \fr{1}{2}[(v_H + R_H)^2 + I_H^2]$.

The mass of xion $ m_{a o }$ in the first line is acquired from mixing of the particle  with $\pi^0$ and $\eta$ pseudoscalars. 
Term in Eq.  \eqref{s183} provides  kinematic term of axion with $f^2_{a}$ is sum over all VEVs of scalars. Beside ordinary dimension 6
terms, there are dimension 5 terms associated with VEVs in the form  $\left(\fr{1}{ f_a}\right)^2 v_H \pa_\mu a \pa^\mu a H $. When $H = \phi$, then we have a  key term proportional to $1/f_a$ only:  $\left(\fr{1}{ f_a}\right)^2 v_\phi \pa_\mu a \pa^\mu a \Phi \sim \left(\fr{1}{ f_a}\right) \, \Phi\,\pa_\mu a \pa^\mu a  $.  The last two terms in the above expression
give couplings of axion with scalar and gauge fields.  The last term in \eqref{s185} contains unwanted mixing of axion with Goldstone bosons eaten by massive $Z$ and $Z^\prime$ bosons. Fortunately, this mixing can be rotated  away by the $U(1)$ rotation.
It is emphasized that the last term in the second line, the right handed Majorana neutrinos are very heavy with mass around
$10^7 \, \gev$. 
   
   The up to $1/f_a$  couplings of  axion field are presented in  Table \ref{tab3}. 
However, the coupling of two axions with inflation is included in bottom of the table.
{\large
		\begin{table}[th]
			\resizebox{10cm}{!}{
				\begin{tabular}{c c c c c c }
					\hline
					&\,  Coupling\,  &\hs  Expression \hs  &\hs  Vertex\hs  & \hs Figure \hs &\hs  Note \\ \hline
					& $a\,F\,F$ & $ \left(\fr{x_f}{ f_a}\right) \pa_\mu \, a \, 
					\bar{f} \ga^\mu  \ga_5 f $ & $ I_{(c,f)} \fr{x_f}{f_a}\slashed{k} \ga_5 $& & $ 2\fr{ 3}{2 f_a } P_R\, \slashed{k}  $ for  $N_R$ \\ \hline
  & $ a\, a\,  R_\phi $& $ \left(\fr{x_\va}{ f_a}\right)^2 \, v_\phi \pa^\mu a \pa_\mu a \, R_\phi$& $ - 2 i  \left(\fr{x_\va}{ f_a}\right)^2 \, v_\phi k.q $ & &$- 2 i  \left(\fr{x_\va^2}{ f_a}\right) \,  k.q $\,  large \\ \hline
			\end{tabular}}
			\caption{ $k,q$ are ingoing  momentum of axion and scalar, $A, S$ and $Z$ are  $CP$-odd,  $CP$-even and
			neural gauge bosons, respectively}
			\label{tab3}
	\end{table}
	
}

\subsection{\label{el} Elimination of mixing between  axion and Goldstone bosons  $G_Z$ and $ G_{Z^\prime}$}
As seen in sub-section \ref{sax}, there are  terms of mixing axion with Goldstone bosons $G_Z$ and $ G_{Z^\prime}$. 
To avoid this trouble \cite{giogi},  we rotate scalar triplets   by opposite $U(1)_{PQ}$ as follows
\bea
\chi & \rightarrow & \chi^\prime = e^{ - i \left(\fr{a}{2 f_a}\right) }\chi \,,\crn
\eta & \rightarrow & \eta^\prime = e^{ - i \left(\fr{a}{2 f_a}\right) }\eta \,,\crn
\rho & \rightarrow & \rho^\prime = e^{  i \left(\fr{a}{2 f_a}\right) }\rho \,.
\label{s241}
\eea
Of course, the quarks have to be changed to keep Yukawa interactions invariant. 
Note that the singlet $\phi$ in unchanged. Then, Eq. \eqref{s185} becomes {\it final} Lagrangian
\bea
\mathcal{L}_a & = & \fr 1 2 \pa_\mu a \pa^\mu a - \fr 1 2 m_{a o }^2 \, a^2 
 +  c_{G G} \fr{\al_s}{4 \pi} \fr{a}{2 f_a} G \tilde{G} + 
c_{WW}\fr{\al_2}{4 \pi} \fr{a}{2 f_a} W \tilde{W}
+ c_{BB}\fr{\al_1}{4 \pi} \fr{a}{2 f_a} B \tilde{B}
\crn
& + & \fr 1 2 \fr{\pa^\mu a}{ f_a} \left(\sum_{f=u,d,T,D}^{l,\nu}  \bar{\psi}_f c_f \ga_\mu \ga_5 \psi  + \fr 1 2  \bar{N}_{a}\, {\bf c}_{N_a} \, \ga^\mu  P_R  N_{a}\right)  -  \left(\bar{q}_L\,  M_a\,  q_R + \mbox{H.c.}\right) \label{s242}\\
& + & \fr 1 2 \, \left(\fr{1}{   f_a}\right)^2  \pa_\mu a \pa^\mu a \left(v_\phi^2 + 2 v_\phi R_\phi + R_\phi^2 \right) 
\,. \label{s243}
\eea 
To have correct axion's  kinetic term, Eq. \eqref{s243} leads to condition
\be
 f_a = v_\phi\,.
\label{s244}
\ee  
The point is worth noting that the terms in last line (doubly ($\pa_\mu a \pa^\mu a$) derivative coupling of axion to inflaton) are characteristic for the model under consideration. Note that this coupling is dimension 5 too. As mentioned in Ref.  \cite{giogi}, the  axion has only two kinds of couplings: derivative couplings to fermions and anomalous
couplings to  gauge bosons $a G \tilde{G}$. It is worth noting that  the new effects mainly happen in the energy  region from $10^7 \, \gev$
to $10^{11} \, \gev$, namely in the region from mass of Majorana right-handed neutrino ($N_R$) to mass of inflaton $\Phi$  .

\subsection{Axion mass from mixing with $\pi^0$}

In the frameworks of chiral perturbative Lagrangian  in the energy scale $ \lesssim 1 \, \gev$ \cite{chp2,chp3},
the axion's  mass was derived \cite{weinberg,stern,luzio,luzio2,giogi,martinez}.
For two quark flavors, the leading order chiral  Lagrangian is given by \cite{luzio}
\bea 
\mathcal{L}^{\chi({ LO})}_{a} &=& 
+\fr{f_\pi^2}{4} \textrm{Tr} \left[(D_\mu U U)^\dagger D_\mu U \right] + \fr{\pa _\mu  a}{2f_a}\fr 1 2 
 \textrm{Tr}\left[c_q \si^a\right] J^{a}_{A,\, \mu} \crn 
 &  & + \fr{f_\pi^2}{2}  B_0 \textrm{Tr} [M_a U^\dag + U M_a^\dag)]   \, .
\label{oct38}
\eea 
The terms in the first line connected with kinetic of fields, while the term in the second line
provides mass of the fields.

Hence we continue with this terms. 
\bea 
 U M_a^\dag  + M_a U^\dag  & = & U M_q +   M_q U^\dag  +   i \fr{a}{2 f_a} \{Q_A,M_q \} U^\dag  
  - 2 \left(\fr{a}{2 f_a}\right)^2\, U (Q_A^2M_q + 2 Q_A M_q Q_A
+ M_q Q_A^2) + \cdots \,,
 \label{oct312}
\eea
where $M_q = \textrm{diag} (m_u,m_d)$.

Using \bea 
U &=& e^{i \fr{\pi^a \si^a}{f_\pi}} = \textbf{1}\,  \cos \left(\fr{\pi}{f_\pi} \right) + i  \fr{\pi^a \si^a}{\pi}\, \sin \left(\fr{\pi}{f_\pi} \right)\,,
 \label{oct310}
\eea
where $\pi = \sqrt{(\pi^0)^2 + 2 \pi^+ \pi^-}$
and $ Q_A = \fr{ M_q^{-1}}{\mbox{Tr}(M_q^{-1})}$, one gets
\bea 
\mathcal{L}_{amass } & = &   \fr{f_\pi^2}{2} B_0 \textrm{Tr} [M_a U^\dag + U M_a^\dag)] = f_\pi^2 m_\pi^2 \cos \left(\fr{\pi}{f_\pi} \right)-\fr{m_a^2}2\,.
\label{oct42}
\eea
Here, the axion mass is given by 
\be
m_a^2    =   \fr{m_\pi^2 f_\pi^2}{f_a^2} \fr{m_u m_d}{(m_u + m_d)^2}\cos \left(\fr{\pi}{f_\pi} \right) \simeq \fr{m_\pi^2 f_\pi^2}{f_a^2} \fr{m_u m_d}{(m_u + m_d)^2}\,.
\label{oct43}
\ee 
A consequence is as follows
\be
m_a \simeq 5.7 \left( \fr{10^{12} \, \gev}{f_a} \right) \, \mu \textrm{eV}\,.
\label{oct46}
\ee

In a similar way, we can   deal with  three flavour quark  (with $s$ quark). The result is given by
\bea
m_a^2  &  = &   \fr{1}{2} m_\pi^2
   \left(\fr{f_\pi}{ f_a}\right)^2 \fr{1}{(m_u + m_d)}\left(\fr{m_u m_d m_s}{m_u  m_d + m_u  m_s +m_d  m_s }\right) \,.
\label{oct7}
\eea
This result  coincides   with one given  in Ref. \cite{sr}
\bea m_a &=& 4  \fr{f_\pi m_\pi}{f_a/N}\left[\fr{ m_u m_d m_s}{(m_u m_d + m_u m_s + m_d m_s)(m_u + m_d)}  \right]^{\fr 1 2}\crn
&\simeq & (1.2 \times 10^{-5} \, \textrm{eV})\left( \fr{10^{12} \, \gev}{f_a/N} \right)\,,
\label{s121}
\eea

\section{\label{conc}Conclusions}

In this paper, we have realized PQ formalism of the 3-3-1 model with Cosmological Inflation. 
The singlet field $\phi$ takes complex
value everywhere, and axion is complex phase in polar coordinates. Then the axion
also appears as a phase of PQ transformations. 
Using the  GKS formation, we have constructed PQ charge operator $Q_A $in terms of diagonal  generators 
 of $SU(3)_L$ subgroup.
The formula shows that the difference of PQ charges ($Q_A$) of up and down quarks equal  to  2, i.e.,
 $\De Q_A =2$, while for electric charge, as usually  $\De Q =1$. For right-handed fermions, 
 electric charges ($Q$) of left-handed and right-handed are assumed to be equal, while for ($Q_A$) is the opposite.
  PQ charges of neutral scalars equal $\pm 2$, while for charged scalars, it vanishes. 
  In the model under consideration, in contract to photon which couples to   charged particles only, while 
   the axion  does not couple to    charged  scalar/gauge bosons.   
To have correct kinetic term for the axion, the PQ scale $f_a$ is to equal  VEV of the singlet scalar boson, namely 
$f_a= v_\phi$. The derivative  couplings of axion to  fermions are presented.   
 The point is worth emphasizing that the axion has doubly derivative coupling with scalar
  playing the role  of inflaton.   This  coupling increases magnitude of coupling to $1/f_a$.  
It is worth emphasizing  that  the new effects mainly happen in the energy  region from $10^7 \, \gev$
to $10^{11} \, \gev$, namely in the region from mass of Majorana right-handed neutrino ($N_R$)
 to mass of inflaton $\Phi$.  
   The chiral effective  Lagrangian as usually provides axion mass consistent with model-independent prediction.

\section*{Acknowledgments}

The authors thank D. T. Nhung for useful discussion.

\appendix

\section{\label{pqo}PQ transformation for chiral fermions}
Proof of  Eq. \eqref{pqr}
\bea
f & \rightarrow & f^\prime =  e^{ i  \left(\fr{x_f}{2 f_a}\right) \ga_5 a} f\,, \hs {\bar f} \rightarrow {\bar f}^\prime =  {\bar f} e^{ i  \left(\fr{x_f}{2 f_a}\right)\ga_5 a}\,, \hs
\va \rightarrow \va^\prime = e^{ i \left(\fr{x_\va}{2 f_a}\right) a} \va \,
\label{pqr1t}\\
 f_L &\rightarrow& f^\prime_L  =  e^{ - i  \left(\fr{x_f}{2 f_a}\right) a} f_L\,,\hs 
{\bar f}_L \rightarrow {\bar f}^\prime_L  = {\bar f}_L  e^{ + i  \left(\fr{x_f}{2 f_a}\right) a} \label{hay1t}
\eea
Noting that
\bea
\ga_5 P_L  & = &P_L\ga_5= - P_L \to (\ga_5)^k P_L =P_L(\ga_5)^k= (- 1)^k P_L \,,\crn
\ga_5 P_R & = &P_R \ga_5 = P_R \to (\ga_5)^k P_R =P_R(\ga_5)^k=  P_R \,.
\label{pqr131}
\eea
As a consequence, we have 
\begin{align}
	 P_Le^{ i  \left(\fr{x_f}{2 f_a}\right)\ga_5 a} =& \left[ \sum_{k=0}^{\infty} \fr{P_L (\ga_5)^k}{k!} \times \left(\fr{ix_fa}{2 f_a}\right)^k\right] = \left[ \sum_{k=0}^{\infty} \fr{(-1)^kP_L}{k!} \times \left(\fr{ix_fa}{2 f_a}\right)^k\right] =e^{ -i  \left(\fr{x_f}{2 f_a}\right) a}P_L=e^{ i  \left(\fr{x_f}{2 f_a}\right)\ga_5 a}P_L
\crn \Rightarrow f_L &\to f'_L=P_L f'=P_L \left(e^{ i  \left(\fr{x_f}{2 f_a}\right)\ga_5 a} f\right)=  e^{ -i  \left(\fr{x_f}{2 f_a}\right)a} P_Lf=  e^{ -i  \left(\fr{x_f}{2 f_a}\right) a} f_L,
\crn P_Re^{ i  \left(\fr{x_f}{2 f_a}\right)\ga_5 a} =&e^{ i  \left(\fr{x_f}{2 f_a}\right)\ga_5 a} P_R = e^{ i  \left(\fr{x_f}{2 f_a}\right) a}P_R \Rightarrow f_R\to f'_R =e^{i  \left(\fr{x_f}{2 f_a}\right) a} f_R. 
\end{align}
The Dirac conjugation:
\begin{align}
	\overline{f} \to & \overline{f'}= \left(f'\right)^{\dagger} \gamma_0=  \left(e^{ i  \left(\fr{x_f}{2 f_a}\right) \ga_5 a}  f\right)^{\dagger} \gamma_0=  f^{\dagger} e^{ -i  \left(\fr{x_f}{2 f_a}\right) \ga_5 a}  \gamma_0= f^{\dagger}   \gamma_0e^{ i  \left(\fr{x_f}{2 f_a}\right) \ga_5 a} =\overline{f} e^{ i  \left(\fr{x_f}{2 f_a}\right) \ga_5 a},
\crn 	\overline{f_L} \to & \overline{f'_L}= \left(e^{ i  \left(\fr{x_f}{2 f_a}\right) \ga_5 a}  f\right)^{\dagger} P_L \gamma_0=   f^{\dagger} e^{ -i  \left(\fr{x_f}{2 f_a}\right) \ga_5 a} P_L \gamma_0=f^{\dagger}  P_L \gamma_0 \times e^{ i  \left(\fr{x_f}{2 f_a}\right) a} =\overline{f_L}e^{ i  \left(\fr{x_f}{2 f_a}\right) a}
\crn 	\overline{f_R} \to & \overline{f'_R}= \left(e^{ i  \left(\fr{x_f}{2 f_a}\right) \ga_5 a}  f\right)^{\dagger} P_R \gamma_0=   f^{\dagger} e^{ -i  \left(\fr{x_f}{2 f_a}\right) \ga_5 a} P_R \gamma_0=f^{\dagger}  P_R \gamma_0 \times e^{ -i  \left(\fr{x_f}{2 f_a}\right) a} =\overline{f_R}e^{ -i  \left(\fr{x_f}{2 f_a}\right) a}. 
\end{align}

This means that in vector fermion, PQ transformation has $\ga_5$, but in chiral form without $\ga_5$.

Therefore, the  mass matrices have  usual form as in ordinary  in the 3-3-1 model \cite{alp331}.

\section{\label{formula}Formula of PQ charge operator}

The PQ charges given in Table \ref{tab2} allows us to write some nice formula as generalized lepton number in 
\cite{cl,jh18}.  Let us write PQ charge operator in diagonal operators as follows (for left-handed fermions sitting in non-singlets).
\be 
Q_A = \al \, T_3 + \bet \, T_8 + \de\,  \mathcal{X}_{pq}\,,
\label{nt1}
\ee
Applying for $Q_3$, one gets
\be \al= + 2, \hs \bet = -\fr 2{\sqrt{3}}\,, \hs \de\, \mathcal{X}_{pq}(Q_3) = + \fr 1 3
\label{nt2}
\ee	
Assuming $\de = 1$, one gets $\mathcal{X}_{pq}(Q_3) = + \fr 1 3$. Hence
\be 
Q_A =  2 \,  T_3 - \fr 2{\sqrt{3}}\,  T_8  +  \mathcal{X}_{pq}\,.
\label{nt3}
\ee
For all fermion triplets, one has $\mathcal{X}_{pq}(\bf 3) = \, \fr 1 3 $.

Note that the above formula is applicable  for left-handed fermions, for right-handed fermions just take opposite  

For scalars
\be 
\mathcal{X}_{pq}(\chi, \eta) =  \fr 4 3\,, \hs \mathcal{X}_{pq}(\rho) = - \fr 2 3\,.
\label{nt4}
\ee 
The PQ charge of the singlet $\phi$ follows from Yukawa coupling and equal  2.

Let us connect PQ charge with electric one being as follows 

\be 
Q =  T_3  - \fr{1}{\sqrt{3}} \, T_8 +  N \,,
\label{nt5}
\ee
From Eqs. \eqref{nt3} and \eqref{nt5}, it follows

\be 
Q_A =  2 \, Q   + \mathcal{X}_{pq}  -  2\,  N \,.
\label{nt6}
\ee
For singlets, their values are given from Yukawa interactions as follows
\be \mathcal{X}_{pq}(N_R) = -1\, ,\hs  \mathcal{X}_{pq}(f_R) = - Q_A(f_L)\,, \hs  \mathcal{X}_{pq}(\phi) = 2\,.
\label{s154}
\ee

For minimal 3-3-1 model, PQ symmetry may be not suitable because of the Landau pole around 5 TeV.
Note that electric charges of up and down elements differ by one unit, while PQ charges of the above elements differ
by two.   That is why the factor $2$ in Eq. \eqref{nt4}.

Therefore we can write PQ transformation for left/right handed fermion and scalar as follows
\bea \Psi(x)& \rightarrow & e^{i\fr{a}{2 f_a} Q_A } \Psi(x)\,, \hs \Psi(x) = f_L, \phi (\mathrm{scalar \, fields})\,,
\label{s151}\\
\overline{\Psi(x)} & \rightarrow & \overline{\Psi(x)} e^{- i\fr{a}{2 f_a} Q_A } \,.
\label{s152}
\eea

\end{document}